# Thermoelectric properties of rare earth filled type-I like Clathrate, $Dy_8Al_{16}Si_{30}$

Kalpna Rajput and Satish Vitta

*Department of Metallurgical Engineering and Materials Science*

*Indian Institute of Technology Bombay, Mumbai 400076, India*

*\* e-mail: satish.vitta@iitb.ac.in & kalpna.rajput@iitb.ac.in*

**ABSTRACT**

Type-I clathrates with a cage structure are known to be of importance for thermoelectric applications as the cage can be filled with a guest atom which leads to reduced thermal conductivity. Among the type-I clathrates, Si-based alloys are of relevance for high temperature application and most importantly because they are made of earth abundant elements. Dysprosium (Dy) has been chosen as the guest atom because of its large mass and small size compared to divalent alkali metal ion. The $Dy_8Al_{16}Si_{30}$ (DAS) alloy has been synthesized by arc melting of pure elements followed by annealing at 780 K for 7 days. Structural characterization performed using XRD and SEM indicates presence of both binary and ternary silicides, $DySi_2$, $DyAl_2Si_2$ together with Al solid solution and Si. The phase mixture remains unchanged even after annealing. The microstructure has a typical dendritic structure with interdendritic phases, signifying a slow, liquid transformation after arc melting. The Seebeck coefficient is found to be positive, a p-type and increases with increasing temperature both before and after annealing. The resistivity is found to be low in the whole temperature range, 2 to 10 µΩm and increases with increasing temperature. The power factor in the as-prepared state is found to be higher at all temperatures in the range 300 K to 700 K compared to annealed state. The thermal conductivity however has been found to decrease on annealing from an unusually high value of ~ 100 $Wm^{-1}K^{-1}$ to 50 $Wm^{-1}K^{-1}$.

**Introduction**

Thermoelectric materials offer a scalable method for converting waste heat into useful electrical energy. This requires an efficient and optimized transport of both charge and heat in the material. Several classes of materials are being investigated for their thermoelectric behavior and intermetallic clathrates are one such class. These compounds have a cage like structure formed by different types of polyhedra with the guest atoms situated inside these cages. Hence they are generally referred to as guest-host cage structures. Among the several types of clathrates, intermetallic type-I clathrate compounds are of significant interest for thermoelectric behavior as they conform to the phonon-glass-electron-crystal concept. These compounds have two distinct types of cages – pentagonal dodecahedron and tetrakaidecahedron formed by the host atoms. The two pentagonal dodecahedron and the six tetrakaidecahedrons that form the unit cell provide 8 sites for the guest atoms which are loosely bound to the cage atoms. These 8 guest atoms are known to result in reducing the total thermal conductivity of the crystalline solid to values close to that of an amorphous solid. Also, the host framework structure offers interchangeability of elements which in turn provides an opportunity to control both the sizes of the cages as well as the electronic properties of the guest-host structure [1-5]. This results in an enhanced thermoelectric dimensionless figure-of-merit ZT given by $ZT = \sigma \alpha^2 T / \kappa$, where α is the Seebeck coefficient, σ the electrical conductivity and κ the total thermal conductivity.

Ternary intermetallic type-I clathrates with the general formula $R_8M_{16}Z_{30}$ where R belongs to divalent alkaline earth metals, M belongs to group 13 semimetal or metal and Z belongs to group 14 elements such as Si, Ge and Sn are extensively synthesized and studied. Among these $Ba_8Ga_{16}Ge_{30}$ compound has been found to exhibit the highest ZT of 1.63 at 1100 K and 1.1 at 900 K for the n-type and p-type variants respectively [6]. However incorporation of a rare earth element into the guest sites has been found to result in enhancing the power factor significantly. Hence the divalent alkaline earth guest metal has been partially as well as completely replaced with divalent rare earth elements such as Eu, Yb in $Ba_8Ga_{16}Ge_{30}$ and it is found that the power factor indeed gets affected [7,8]. Although these clathrates exhibit a high ZT at high temperatures, they consist of low melting, heavy, non-earth abundant host elements such as Ga and Ge. Hence in the present work these are completely substituted with Al and Si which are non-toxic, earth abundant low cost elements. Since Si and Al have higher melting temperatures compared to Ga and Ge, they will have higher thermal stability at high temperatures. These host elements are also higher compares to Ga and Ge which will result in lower density for the devices which is an important technological consideration. The divalent guest element has been completely substituted with trivalent rare earth element in the present work. Dy has been chosen in the present work as its size is smaller than Ba and also heavier than Ba which should result in significant reduction in total thermal conductivity. Earlier studies have found that the rare earth elements with their 4f electrons having a high tendency to hybridize with the host lattice conduction electrons, are difficult to be introduced in large amounts [9]. Hence the current work is an attempt in this direction.

**Experimental Methods**

The intermetallic clathrate compound, $Dy_8Al_{16}Si_{30}$ was synthesized by arc melting in an argon atmosphere. High purity, 99.9% elements Dy, Al and Si in the atomic ratio 8:16:30 were used to synthesize the stoichiometric compound. The melting process was repeated 3 times to obtain good homogeneity by flipping the ingot between each melt process. The ingot was cut into a rectangular bar of size (10×8×6) mm$^3$ for thermophysical properties characterization. The bar was annealed at 780 K for 7 days to further homogenize and nucleate the type-I phase. The thermophysical properties of annealed bar sample were also measured and compared with the as-prepared sample. The density of the bar sample determined using Archimedes principle was found to be 3.9 g cm$^{-3}$, close to the theoretical density of the clathrate compound.

The microstructural characterization was performed by X-ray diffraction using Cu-Kα radiation in a Panalytical X'pert Pro diffractometer as well as using a scanning electron microscope equipped with energy dispersive X-ray spectroscopy detector. For X-ray diffraction the sample was powdered in a mortar with a pestle. Scanning electron microscopy was performed on a polished and etched sample surface. The heat capacity $C_P$ of as-prepared and annealed samples was measured using a differential scanning calorimeter, Netzsch DSC 204 F1 phoenix under continuous nitrogen flow at a heating rate of 5 K min$^{-1}$. The Seebeck Coefficient as well as the electrical resistivity were measured using ULVAC ZEM-3 in a helium atmosphere in the temperature range 300 K to 700 K. The thermal diffusivity was determined using the Laser Flash System LFA 457 in nitrogen atmosphere. The total thermal conductivity was calculated using the experimentally determined $C_p$, density and diffusivity.

**Results and Discussion**

The microstructure of the polished and etched sample as seen in the scanning electron microscope is shown in Figure 1. The dense microstructure does not show any porosity. The microstructure shows a typical cast structure with dendritic and inter-dendritic phases. Such structures form as a result of slow cooling from the liquid phase and are often associated with chemical segregation which leads to multiple phase formation. In order to determine the chemical composition of different phases, a point analysis was performed at different locations as shown in Figure 1(b) and to determine the overall chemical composition, an area analysis was performed as shown in Figure 1(c). All the results are given in table 1. The chemical composition shows the presence of 4 clearly different phases- a solid solution of Si in Al (point 1), a ternary phase similar to $DyAl_2Si_2$ (point 2), pure Si (point 3) and binary intermetallic compound $DySi_2$ (point 4) with small amount of Al. The overall composition indicated a slight loss of the guest element, Dy. The X-ray diffraction pattern in as-prepared and annealed conditions are nearly identical except for the peak intensities as shown in Figure 2. The compound $Dy_8Al_{16}Si_{30}$ has not been reported to be synthesized with type-I cage structure and this work is first such attempt. Hence the diffraction patterns were compared to the structure of compositionally similar compounds such as $Ba_8Al_{16}Si_{30}$ and $Ba_8Ga_{16}Ge_{30}$. The diffraction patterns do not exhibit peaks corresponding to type-I clathrate structure, in agreement with the chemical compositions determined in the scanning electron microscope. The diffraction patterns exhibit peaks corresponding to a mixture of phases such as –Si, solid solution of Si in Al, binary intermetallic $DySi_2$ and the ternary compound $DyAl_2Si_2$. The ternary compound has a trigonal P-3m1 structure while the binary intermetallic compound has a tetragonal $I4_1/amd$ structure. These results clearly show that the ternary type-I clathrate structure is not present both in as-prepared and annealed conditions. In earlier studies also type-I clathrates both with and without rare-earth guest atoms, it is found that stabilizing a single phase is extremely difficult and secondary phases are always present together with the type-I clathrate phase in varying fractions [8,10,11,12]. The incorporation of trivalent rare-earth guest atoms in stoichiometric quantities is reported to be extremely difficult [9, 13]. The specific reasons for this behavior however are not clearly known at present.

The variation of the heat capacity, $C_p$ as a function of temperature is shown in Figure 3 for the as-prepared sample. It shows two clear endothermic peaks at ~ 850 K and 923 K, indicating a transformation at these two temperatures. The phase that can undergo a phase transformation at these relatively low temperatures is Al-solid solution and Si. The Al-Si phase diagram shows that Al-solid solution and Si react (eutectic) at ~ 850 K to form a partial liquid phase and undergoes complete melting at 923 K [14]. Hence it can be concluded that the two endothermic peaks observed correspond to eutectic reaction between Al and Si and subsequent melting of Al solid solution respectively. Since the annealed microstructure also shows the same phases mixture as in the case of as-prepared alloy, the thermoelectric transport properties have been investigated only till 700 K and are discussed below [14].

The temperature dependent variation of Seebeck coefficient α, resistivity ρ, thermal conductivity κ and power factor, $α^2σ$, of as-prepared and annealed alloy are shown in Figure 4. The resistivity in both conditions is low, varying between 2 and 10 µΩm and it increases slightly with increasing temperature. The resistivity in as-prepared condition however is lower compared to annealed state of the alloy. The absolute values of resistivity as well as their temperature dependence indicates a metallic behavior in both cases. The temperature dependence of conductivity is found to obey approximately a $T^{-1/2}$ behavior, indicative of disorder dominated scattering [11]. However, since the alloys have multiple phases it is not possible to identify a single scattering mechanism contributing to overall electrical conductivity. The Seebeck Coefficient or thermopower, α is found to be positive in the entire temperature range indicating a p-type behavior contrary to generally observed n-type behavior [11,12,15]. The absolute value of α however is found to be small, varying between 5 and 20 µVK$^{-1}$ with a positive temperature dependence. The small values of α combined with low electrical resistivity indicate a metallic nature with a large effective carrier concentration. This behavior is indeed reflected in the total thermal conductivity, κ determined from the experimentally determined heat capacity, thermal diffusivity and density and shown in Figure 4(c). The thermal conductivity is extremely large and is dominated by the charge carriers. The thermal conductivity of as-prepared alloy decreases with increasing temperature while that of annealed alloy increases with increasing temperature and then decreases. It is to be noted that annealed alloy exhibits a small peak like feature in the temperature dependence of α, ρ and κ at ~ 550 K. The heat capacity however does not show any anomaly at this temperature. This is probably due to a change in the effective transport mechanism which is not accompanied by any structural transitions. Since the alloy in the annealed condition has multiple phases, it is not possible to attribute this to any mechanism or phase at present. An estimation of the power factor, $α^2σ$ in the two states shown in Figure 4(d) clearly shows that annealing leads to a deterioration mainly due to resistivity increase. The thermal conductivity of the annealed alloy however is lower, but insufficient to offset the advantage of higher power factor of the as-prepared alloy.

**Conclusions**

A type-I clathrate compound corresponding to $Dy_8Al_{16}Si_{30}$ has been attempted for synthesis. It is found that type-I cage compound does not form under the synthesis conditions used presently. The alloy exhibits multiple phases, pure Si to a ternary compound, $DyAl_2Si_2$ which do not mix to form the caged compound $Dy_8Al_{16}Si_{30}$ even after annealing at 780 K for 7 days, clearly indicating that a much higher temperature homogenization is required to nucleate and stabilize the caged compound. These results are in agreement with earlier efforts wherein it was found that complete substitution of the divalent alkali metal with a trivalent rare-earth element is extremely difficult.

**Acknowledgements:** The authors wish to acknowledge Nanomission, Department of Science and Technology, Government of India for financial assistance.

**Figure Captions:**

**Fig. 1** The X-ray powder diffraction pattern of as-prepared (a) and annealed (b) $Dy_8Al_{16}Si_{30}$ indicate the presence of multiple phases in both cases. The different phases that have been identified are marked and they vary from Si to the ternary compound $DyAl_2Si_2$.

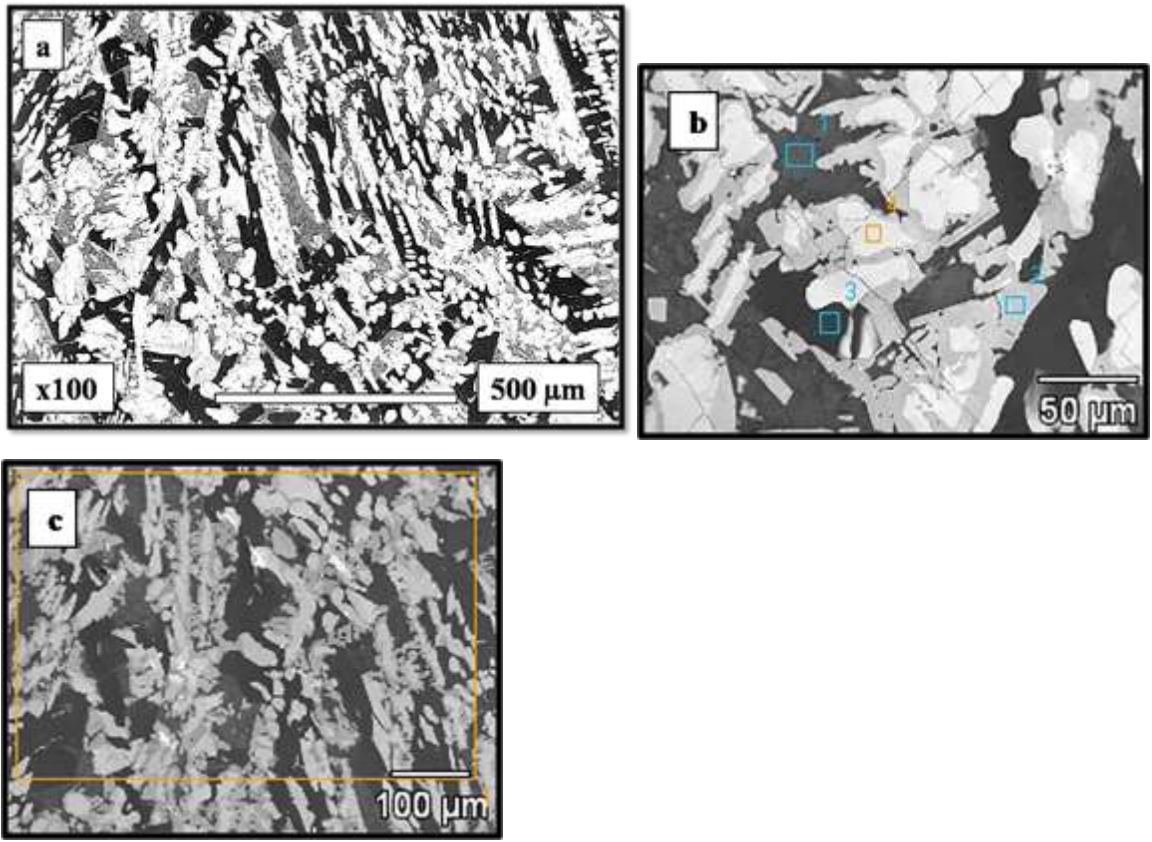

**Fig. 2** The microstructure of polished and etched as-prepared Dy$_8$Al$_{16}$Si$_{30}$ alloy shows a dendritic structure with interdendritic phases, (a). Chemical composition determined at different regions, (b) indicates different phases compared to the overall chemical composition of the area as shown in (c).

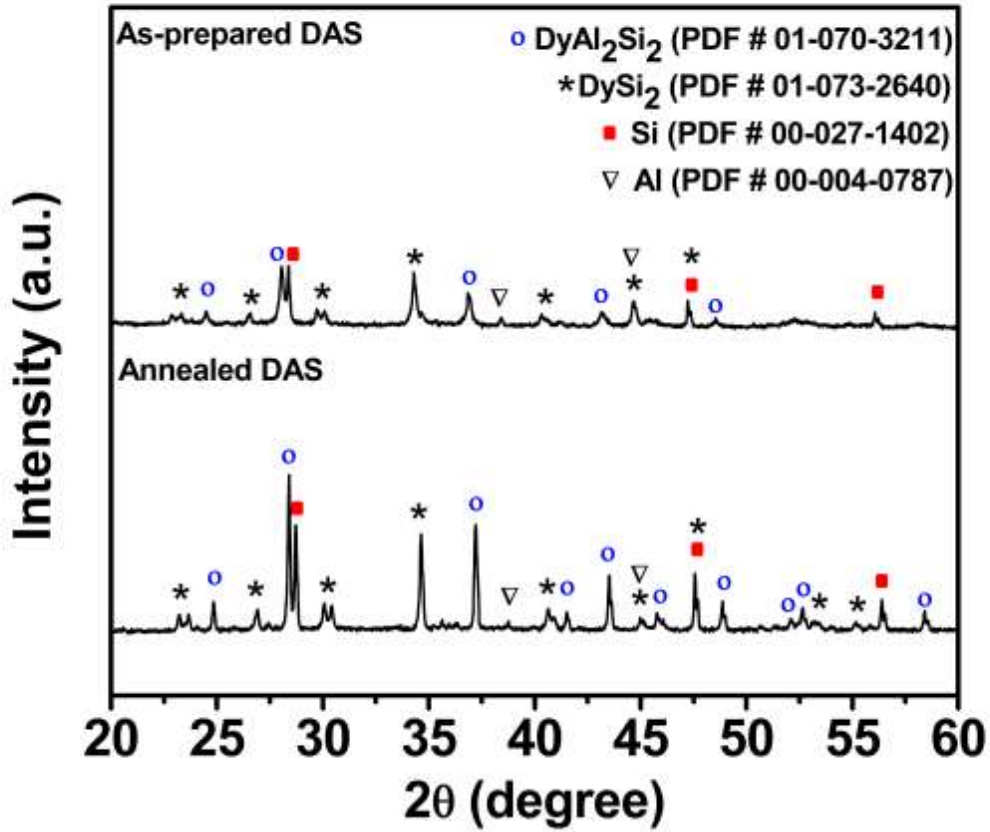

**Fig. 3** The heat capacity $C_p$ determined as a function of temperature of the as-prepared alloy shows two clear endothermic peaks at 850 K and 923 K. These peaks indicate transformation of phases in the alloy.

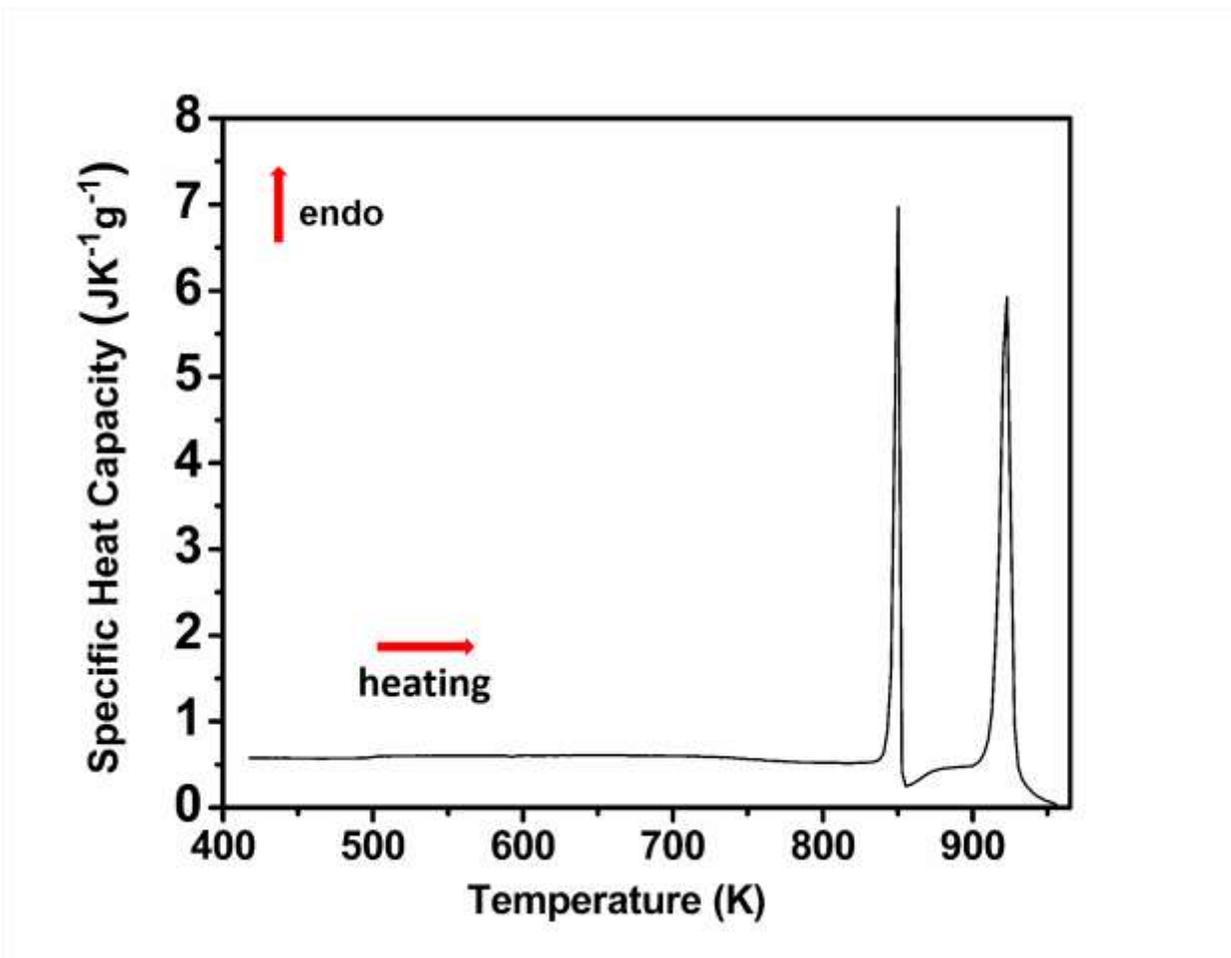

**Fig. 4** The thermoelectric properties, (a) Seebeck Coefficient α, (b) resistivity ρ, (c) total thermal conductivity κ, and (d) power factor α²σ of both as-prepared and annealed alloys are shown as a function of temperature. The Seebeck Coefficient α in both the cases is positive indicating a p-type behavior.

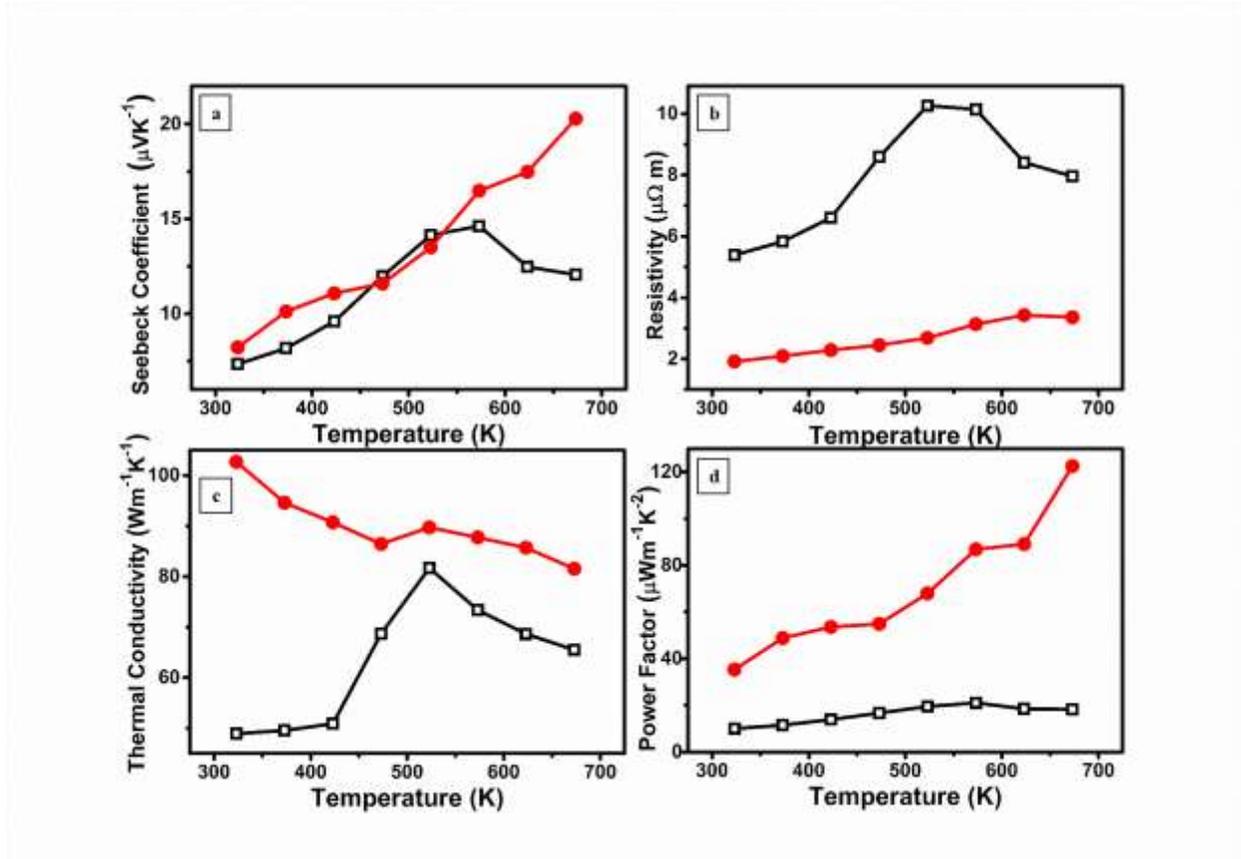

**Table 1** The chemical composition of different constituents of the microstructure as determined by energy dispersive x-ray analysis in scanning electron microscope are given here. The points refer to different locations as seen in the SEM micrograph, Figure 1(b) and the area, as shown in Figure 1(c). The expected overall chemical composition of the alloy is given in brackets.

| At.% | Dy | Al | Si |
|---|---|---|---|
| **Point 1** | 0.15 | 96.1 | 3.7 |
| **Point 2** | 22.3 | 40.2 | 37.5 |
| **Point 3** | | | 100 |
| **Point 4** | 38.7 | 4.2 | 57.1 |
| **Area** | 9 (14.8) | 30.3 (29.6) | 57.6 (55.6) |